\documentclass[%
 reprint,
nobibnotes,
 amsmath,amssymb,
 aps,
 prb,
]{revtex4-1}

\usepackage{docs}%
\usepackage{siunitx}

\DeclareSIUnit[number-unit-product = {\,}]
\cal{cal}
\DeclareSIUnit\kcal{\kilo\cal}
\DeclareSIUnit[number-unit-product = {\,}]
\Btu{Btu}
\DeclareSIUnit[number-unit-product = {\,}]
\Fahr{\degree F}
\DeclareSIUnit[number-unit-product = {\,}]
\usepackage{bm}%
\usepackage[colorlinks=true,linkcolor=blue]{hyperref}%
\usepackage{graphicx}
\usepackage[utf8]{inputenc}
\usepackage{tabularx} % provides a column type called "X" that should satisfy your professed need to have several equal-width columns

\usepackage{dcolumn}
\usepackage{epstopdf}
\usepackage{afterpage}
\usepackage{setspace}
\usepackage{multirow}
\usepackage{dsfont}

\expandafter\ifx\csname package@font\endcsname\relax\else
 \expandafter\expandafter
 \expandafter\usepackage
 \expandafter\expandafter
 \expandafter{\csname package@font\endcsname}
\fi
\begin{document}
\setstretch{1.0}

\title[]{
    Machine Learning of Free Energies in Chemical Compound Space Using Ensemble Representations:
    Reaching Experimental Uncertainty for Solvation
}

\author{Jan Weinreich}

\affiliation{University of Vienna, Faculty of Physics, Kolingasse 14-16, AT-1090 Wien, Austria}

\author{Nicholas J. Browning}

\affiliation{Institute of Physical Chemistry and National Center for Computational Design and Discovery of Novel Materials (MARVEL), Department of Chemistry, University of Basel, Klingelbergstrasse 80, CH-4056 Basel, Switzerland}

\author{O. Anatole von Lilienfeld}
\email{anatole.vonlilienfeld@univie.ac.at}
\affiliation{University of Vienna, Faculty of Physics, Kolingasse 14-16, AT-1090 Wien, Austria}
\affiliation{Institute of Physical Chemistry and National Center for Computational Design and Discovery of Novel Materials (MARVEL), Department of Chemistry, University of Basel, Klingelbergstrasse 80, CH-4056 Basel, Switzerland}

\date{\today}

\begin{abstract}

    Free energies govern the behavior of soft and liquid matter, and improving their predictions could have a large impact on the development of drugs, electrolytes or homogeneous catalysts.
    Unfortunately, it is challenging to devise an accurate description of effects governing solvation such as hydrogen-bonding, van der Waals interactions, or conformational sampling.
    We present a Free energy Machine Learning (FML) model applicable throughout chemical compound space and based on a representation that employs Boltzmann averages to account for an approximated sampling of configurational space.
    Using the FreeSolv database, FML's out-of-sample prediction errors of experimental hydration free energies decay systematically with training set size, and experimental uncertainty (0.6 kcal/mol) is reached after training on 490 molecules (80\% of FreeSolv).

    Corresponding FML model errors are also on par with state-of-the art physics based approaches.
    To generate the input representation for a new query compound, FML requires approximate and short molecular dynamics runs.
    We showcase its usefulness through analysis of FML solvation free energies for 116k organic molecules (all force-field compatible molecules in QM9 database) identifying the
    most and least solvated systems, and rediscovering  quasi-linear  structure property relationships
    in terms of simple descriptors such as hydrogen-bond donors, number of NH or OH groups, number of oxygen atoms in hydrocarbons, and number of heavy atoms.

    FML's accuracy is maximal when the temperature used for the molecular dynamics simulation to generate averaged input representation samples in training is the same as for the query compounds.
    The sampling time for the representation converges rapidly with respect to the prediction error.

\end{abstract}

\maketitle

\bigskip

\section{Introduction} \label{sec:introduction}

An accurate description of solvation free energy is fundamentally important to rationalizing reaction kinetics and product propensities. Therefore accurate models describing solvation have far reaching utility from drug design to battery development. Computational methods for predicting solvation free energies based on \textit{ab initio} methods\cite{PhysRevB.101.060201,keith2, qmmm, doi:10.1063/1.5089199}, while accurate in principle, impose a substantial computational burden, and are therefore inherently limited when it comes to navigating chemical compound space (CCS). Conversely, more readily available methods based on parametrized force-fields (FFs), implicit solvent models (PCM\cite{poissonB, pcm}, GBSA\cite{TruhlarSMGB}, SMD\cite{smd}, COSMO\cite{cosmo, cosmo2}, COSMO-RS\cite{KLAMT200043}) or hybrid models (3D-RISM\cite{doi:10.1021/jp971083h, KOVALENKO1998237, validationset}) trade reduced computational expense for lower accuracy w.r.t. experiment. In particular, continuum solvation models exhibit several disadvantages including lack of locality in distinct atomic environments\cite{keith1, doi:10.1021/ct4004433}, poor modeling of hydrogen bonding, inaccurate estimates of entropy contributions\cite{doi:10.1021/acs.jctc.7b00169}, as well as poor decoupling between short-range and long-range effects. Still in particular alchemical FF based approaches have become a routine method for free energy calculations\cite{mey2020best}.

The recent success of quantum machine learning (QML) in the domain of theoretical and computational chemistry due to unprecedented availability of calculated single-point geometry quantum data, has been manifested for challenging molecular problems, such as accurate prediction of molecular electronic properties like atomization energies\cite{coulomb, doi:10.1021/acs.jctc.7b00577}, application to elpasolites\cite{PhysRevLett.117.135502}, excited states\cite{westermayr2020machine} or fragment based learning with AMONS\cite{superbing}.

Starting with the work of Behler and Parrinello on high-dimensional neural network potentials\cite{PhysRevLett.98.146401, doi:10.1063/1.4966192} there has been growing interest in applying ML to MD simulations\cite{unke2020machine, doi:10.1021/acs.chemrev.0c00665} e.g. using gradient domain ML\cite{sauceda2020molecular} and molecules in complex environments such as surfaces, water \cite{doi:10.1063/5.0014876,doi:10.1021/acs.jpclett.7b00358, doi:10.1021/acs.jpcb.7b01490, Cheng1110}
or systems under extreme pressure and temperature \cite{10.1038/s41586-020-2677-y}. A promising development are coarse-grained ML models\cite{doi:10.1063/5.0007276, NOE202077,doi:10.1021/acs.jctc.9b01256, doi:10.1063/1.5119101, PhysRevE.100.033302}. Ref.~\cite{huang2020ab} provides a general overview about the most recent developments of ML in CCS.

\begin{figure}
    \centering
    \includegraphics[width=\columnwidth]{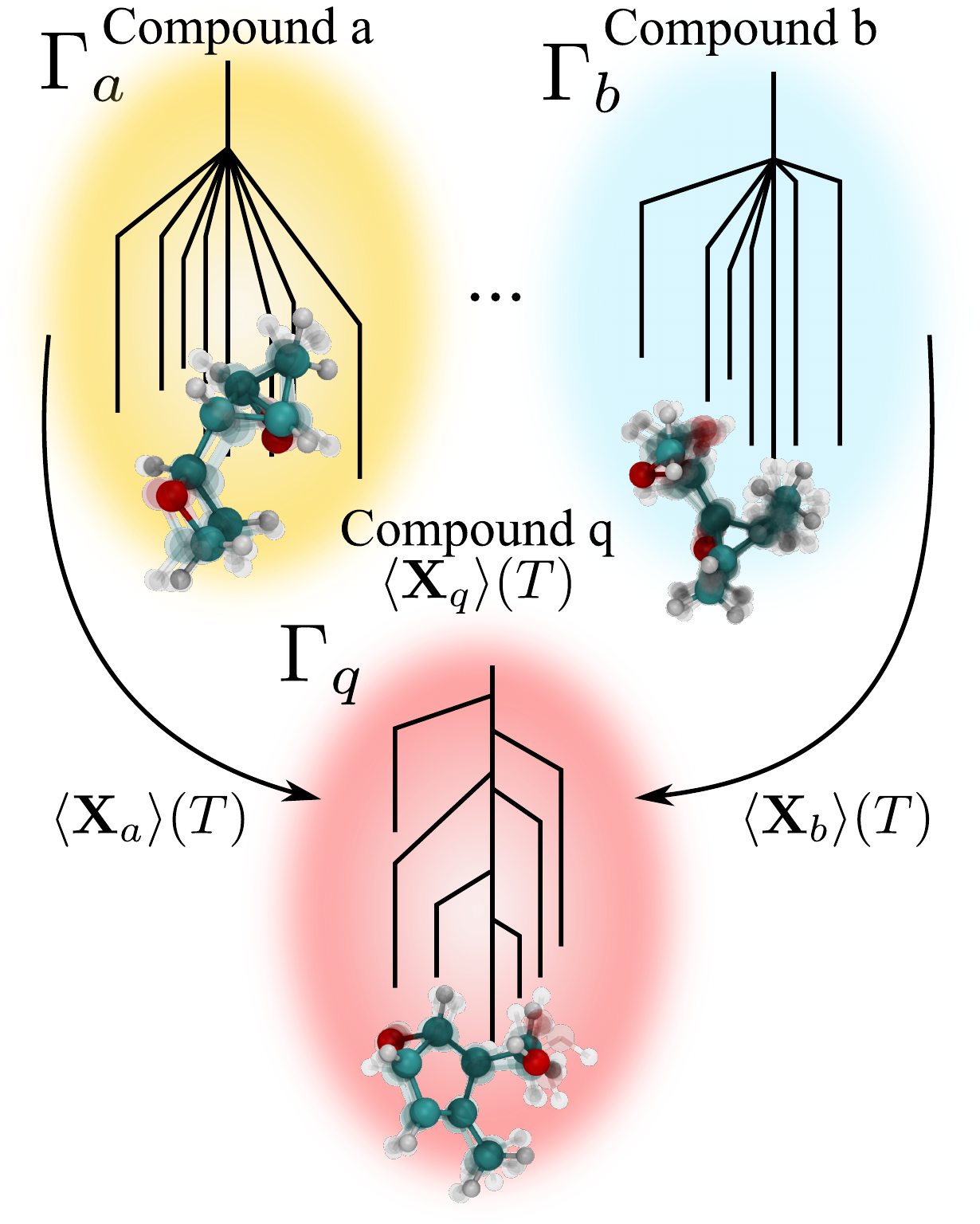}
    \caption{
        The FML prediction for a new query compound $q$ depends on the phase space $\{ \mathbf{\Gamma}_{i} \}$ of all training compounds consisting of conformers, depicted by disconnectivity graphs, sampled at the same temperature $T$, by virtue of ensemble average representations $\langle \mathbf{X} \rangle(T)$.
    }
    \label{fig:sktech_fml}
\end{figure}

ML based free energy models, on the other hand, are much less established, and potential applications to explore CCS in terms of thermodynamic properties have largely remained unexplored except for some very recent publications\cite{doi:10.1021/acs.jcim.0c00479, hybridML, lim2020mlsolva,axelrod2020molecular,vermeire2020transfer}.
Here, we introduce a new ML model capable of predicting ensemble averages, such as free energies of solvation $\Delta G_{\text{sol}}$.
In particular, our free energy ML (FML) model is designed to deliver both, computational efficiency as well as prediction errors which systematically improve with training set size, thereby being able to reach experimental uncertainty levels.
The FML model fills, to the best of our knowledge, an important gap in the field of ML for atomistic simulation by explicitly accounting for an ensemble of molecular conformations through Boltzmann averaged representations\cite{felixFCHL, FCHL,Huang2018}, rather than through fixed geometry based representations.
FML thus avoids the pitfall of neglecting the variance of ensemble properties when basing predictions on fixed geometries only.

As illustrated in Fig.~\ref{fig:sktech_fml} the prediction of an ensemble property for a query compound $q$ at temperature $T$ depends on the phase spaces $\{ \mathbf{\Gamma}_{i} \}$ of all training compounds $i$. From this point of view FML infers predictions by combining the information of all the phase spaces at given temperature $T$ by virtue of the average representation $\langle \mathbf{X} \rangle$, i.e.~as an integral over the configurationally sampled space.
\\

Our paper is structured as follows: We begin by detailing the FML workflow in sec.~\ref{sec:krr_theory} with emphasis on the ensemble based representation. Next we present the results in sec.~\ref{sec:results} starting with a numerical demonstration of the necessity of the ensemble representation before assessing the accuracy of FML. Finally, we demonstrate the feasibility of the method for high throughput free energy predictions of 116k organic molecules (a subset of QM9~\cite{qm9}) revealing trends between molecular structure and solubility.

\section{Theory} \label{sec:theory}

We employ a representation based on an ensemble of conformers generated through MD sampling. This gives rise to a unique and temperature dependent representation of the system state. The resulting machine learning framework FML constitutes a physics based approach, since this representation is rooted in statistical mechanics.
As shown and discussed below a comparison with state-of-the art solvation models reveals that FML retains the promise of being faster, more transferable and extendable than solvation methods based on conventional fitting of model parameters. While we focus on free energies of solvation we note that the same methodology might open new pathways for ML applications to other ensemble properties such as protein binding or enthalpy and entropy.

\subsection{Kernel Ridge Regression} \label{sec:krr_theory}

We use kernel ridge regression\cite{Huang2018} (KRR) a supervised ML method\cite{Vapnik1998}, which can be derived from Gaussian process regression. KRR non-linearly maps input into a high-dimensional feature space which renders the regression problem linear. The similarity between compounds $i$ and $j$ with representations $\mathbf{X}_i, \mathbf{X}_j$ is measured by applying a Gaussian kernel function $K$,
\begin{align}
    K(\mathbf{X}_i, \mathbf{X}_j) = \exp{\left( -\frac{\vert \vert \mathbf{X}_i - \mathbf{X}_j \vert \vert^{2}_{2}}{2 \sigma^2} \right)},
    \label{eq:gauss}
\end{align}
where $\sigma$ is the kernel-width hyperparameter. The prediction of property $p$ of query compound $q$ is given by,
\begin{align}
    p(\mathbf{X}_{q}) = \sum_{i}^{N^{\text{train}}} \alpha_i K(\mathbf{X}_{i}^{\text{train}},\mathbf{X}_{q})~,
    \label{eq:mapping}
\end{align}
where $K(\mathbf{X}_{i}^{\text{train}},\mathbf{X}_q)$ is the kernel function, evaluated on the query and all training compounds with weight coefficients $\boldsymbol{\alpha}$. The unique solution vector for the optimal set of regression coefficients $\boldsymbol{\alpha}$ is given by,
\begin{align}
    \boldsymbol{\alpha} =  (\mathbf{K} + \lambda \cdot \mathds{I})^{-1} \mathbf{p},
    \label{eq:solution}
\end{align}
with the vector $\mathbf{p}$ containing all values of the target property in the training set and regularization parameter $\lambda$.

\subsection{Ensemble Based Representation}

While quantum machine learning (QML) is commonly used as a surrogate model for approximate solutions to the electronic Schr\"odinger equation, i.e.~they are associated with exactly one fixed configuration of atoms in a compound (single point).
By contrast, the free energy is a property of an ensemble of possible configurational states.

\begin{figure}
    \centering
    \includegraphics[width=\columnwidth]{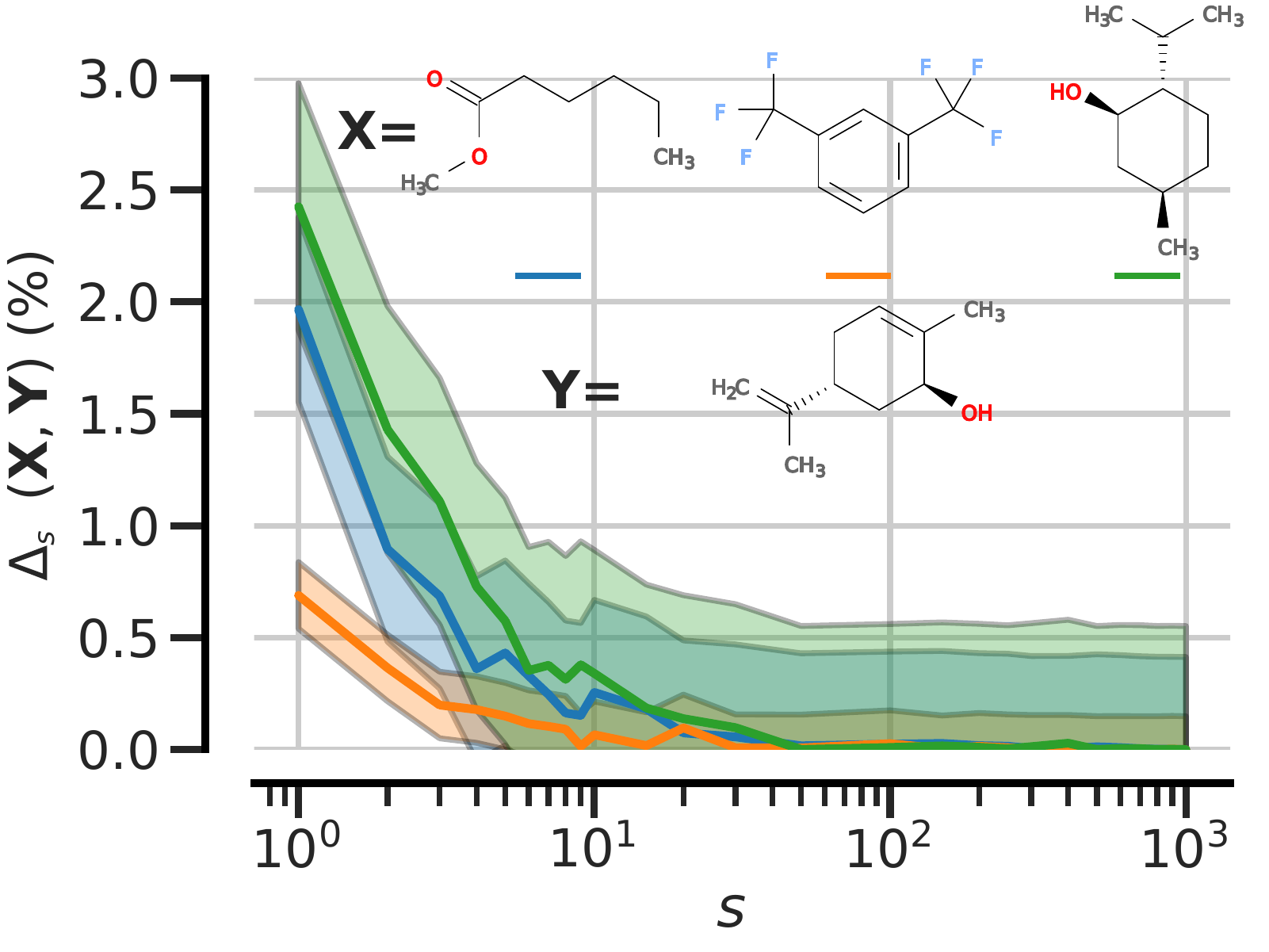}
    \caption{
        Relative deviation $\Delta_{s} (\mathbf{X}, \mathbf{Y})$ w.r.t. taking all 1000 MD samples in percent (s. Eq.~\ref{eq:convergence_norm}) of the distance between the FML representations (s. Eq.~\ref{eq:average_representation}) of three molecules $\mathbf{X}$ and fixed $\mathbf{Y}$ (shown as inset) as a function of the number of MD samples $s$.
    }
    \label{fig:convergence_rep}
\end{figure}
\begin{figure}
    \centering
    \includegraphics[width=\columnwidth]{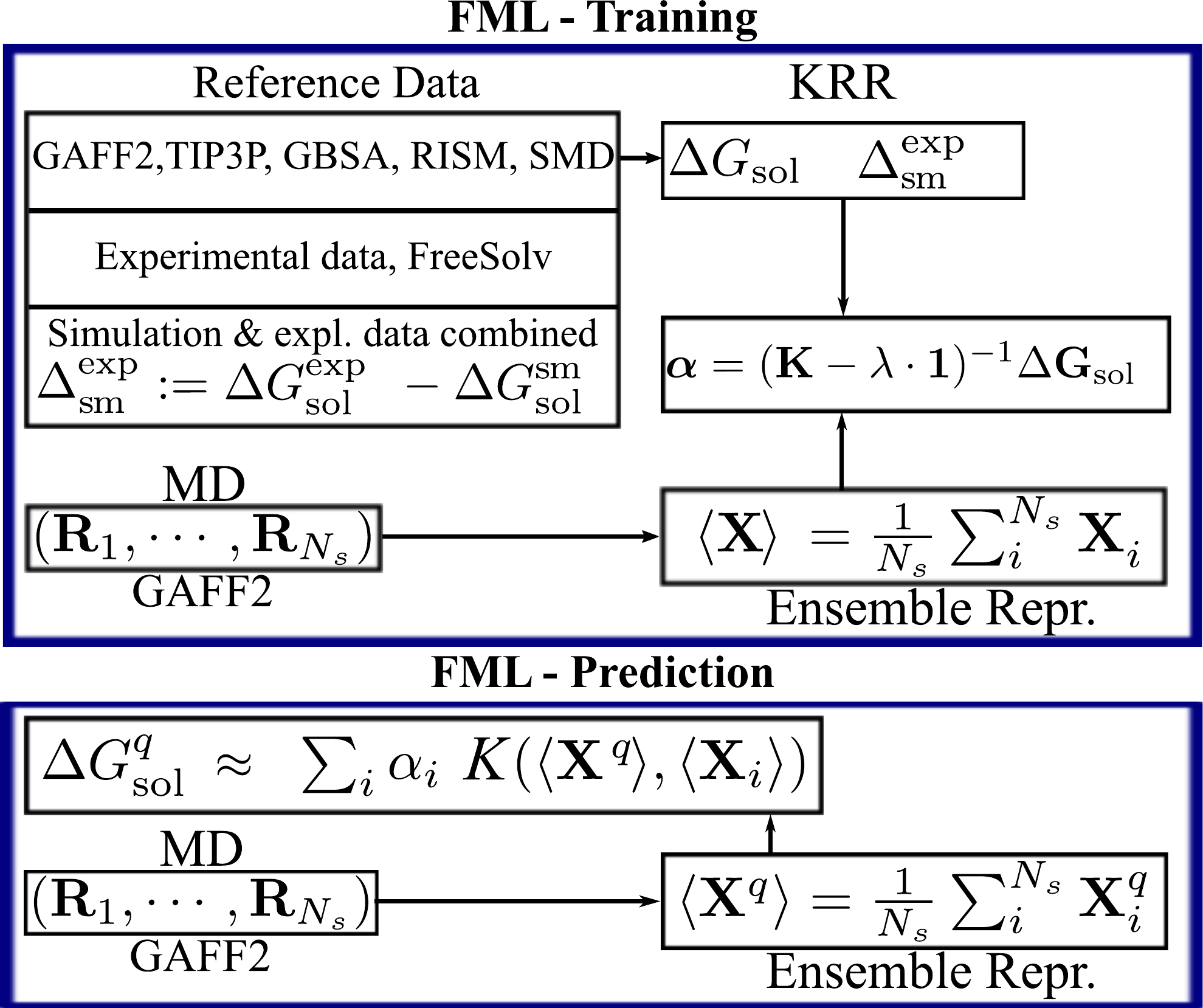}
    \caption{
        Steps for training an FML model for the free energy of solvation $\Delta G_{\text{sol}}$ (upper box) and for prediction of $\Delta G_{_{\text{sol}}}^q$ for a new query compound $q$ (lower box).}
    \label{fig:workflowA}
\end{figure}
Hence, the intriguing question is how to define a physics based representation of a compound for an ensemble property, such as the free energy of solvation.
Here, we have used averages of FCHL19\cite{FCHL, felixFCHL}, a geometry dependent many-body representation that includes two- and three-body terms where the first term accounts for interatomic distances and the second term for relative orientations of triplets of atoms. In this spirit, representations for thermodynamic properties should be designed similarly, taking into account the ensemble of accessible configurations at a given temperature. We used a thermodynamic ensemble average allowing for a unique definition of a representation given a set of configurational snapshots obtained by short MD simulations at temperature $T$ (s. sec.~\ref{sec:comput}). First the FCHL19 representation $\mathbf{X}(\{ \mathbf{r}_i \})$ is computed for all snapshots $i$ before calculating the ensemble average by numerical integration as follows,
\begin{equation}
    \label{eq:average_representation}
    \begin{split}
        \langle \mathbf{X} \rangle (T) &=
        \frac{1}{Z} ~ \int_{\mathbf{\Gamma}(T)}  \mathbf{X}(\{ \mathbf{r}_i \}) e^{-\beta E_i} ~d \mathbf{\Gamma} \\
        & \approx \frac{1}{s} \sum_{i}^{s} \mathbf{X}_{i}~,
    \end{split}
\end{equation}
with $s$ uncorrelated gas-phase solute samples weighted by the respective Boltzmann factor $e^{-\beta E_i}$ with $\beta = \frac{1}{k_{B} T}$ and $Z$ is the partition function. Note that the average representation $\langle \mathbf{X} \rangle (T) $ depends on $T$ since both the integration domain $\mathbf{\Gamma}(T)$ over phase space and the Boltzmann factors depend on $T$.

In Fig.~\ref{fig:convergence_rep} we show the convergence of the pairwise distance $\Delta_{s}$ between average representations $\langle \mathbf{X}_{s}  \rangle $ and $\langle \mathbf{Y}_{s}  \rangle $ for three molecules $\mathbf{X}$ w.r.t. a fixed molecule $\mathbf{Y}$ for a given number of MD samples $s$ which we define as,
\begin{equation}
    \Delta_{s} (\mathbf{X}, \mathbf{Y}) :=  \\
    \left \vert  \frac{ || \langle \mathbf{X}_{s}  \rangle   - \langle \mathbf{Y}_{s} \rangle    || - || \langle \mathbf{X}_{s_{\text{max}}}  \rangle   - \langle \mathbf{Y}_{s_{\text{max}}}  \rangle    || }{|| \langle \mathbf{X}_{s_{\text{max}}}  \rangle   - \langle \mathbf{Y}_{s_{\text{max}}}  \rangle    ||} \right \vert ~,
    \label{eq:convergence_norm}
\end{equation}
with the maximal number of MD samples $s_{\text{max}} = 1000$ and the euclidean norm $||.||$.
For the randomly selected examples shown in Fig.~\ref{fig:convergence_rep} we find that for $s= 10$ uncorrelated molecular geometries the relative deviation is below $0.4~\%$.
Generally, we find that approximately $s= 10$ uncorrelated molecular geometries (from MD s. sec.~\ref{sec:comput}) are sufficient to converge the distances below $4\%$ relative deviation (s. also SI Fig.~4).

Our representation, which we also refer to as the FML representation, is similar to what was recently presented in\cite{doi:10.1063/5.0012230}, however, here we rather use experimental free energies for training because of the inherent inaccuracy of conventional solvation models\cite{doi:10.1021/acs.jctc.7b00169} employed in that work. In addition there are also a number of quantitative structure–activity relationship (QSAR) based approaches for representing conformers\cite{doi:10.1021/ja9718937,qsar2,qsar}.

In the following we refer to KRR using a given vacuum geometry as QML and to free energy machine learning using the ensemble averaged FCHL19 representation as FML. Note that unlike most applications of QML we train on experimental data rather than on computational results from solutions to the Schrödinger Equation.
The methodology of FML is fully transferable to free energies from any level of theory, and it is both atomistic and {\em ab initio} in the sense that solely atomic configurations are required as an input without need for molecular connectivities. The generation of the snapshots via molecular dynamics can be performed using either {\em ab initio}, or force-field based methods just as well.
In Fig.~\ref{fig:workflowA} we show the practical steps for training a FML model for free energies of solvation. First, a set of free energies is calculated, or experimental values are collected from the literature. Subsequently, and for all solute configurations the average over all representations of the trajectories is computed (s. Eq.~\ref{eq:average_representation}). Finally, the regression weights $\boldsymbol{\alpha}$ are determined (s. Eq.~\ref{eq:solution}).

\subsection{Uniqueness} \label{sec:unique_predictions}

To illustrate the importance of an ensemble based approach we use QML to predict a hypothetical value of $\Delta G_{\text{sol}}$ for all possible fixed conformers of several stereo-isomers. We emphasize that for an ML model to be consistent the predictions for free energies of solvation of a set of conformers should differ less than the experimental uncertainty. However, in the following we will show that QML based on a single molecular conformation can indeed lead to inconsistent predictions. For training the QML model we use experimental free energies and representation vectors $\mathbf{X}_{\text{vac}}$ based on vacuum geometries provided by the FreeSolv\cite{FreeSolv} database.

\begin{figure}[htb]
    \centering
    \includegraphics[width=\columnwidth]{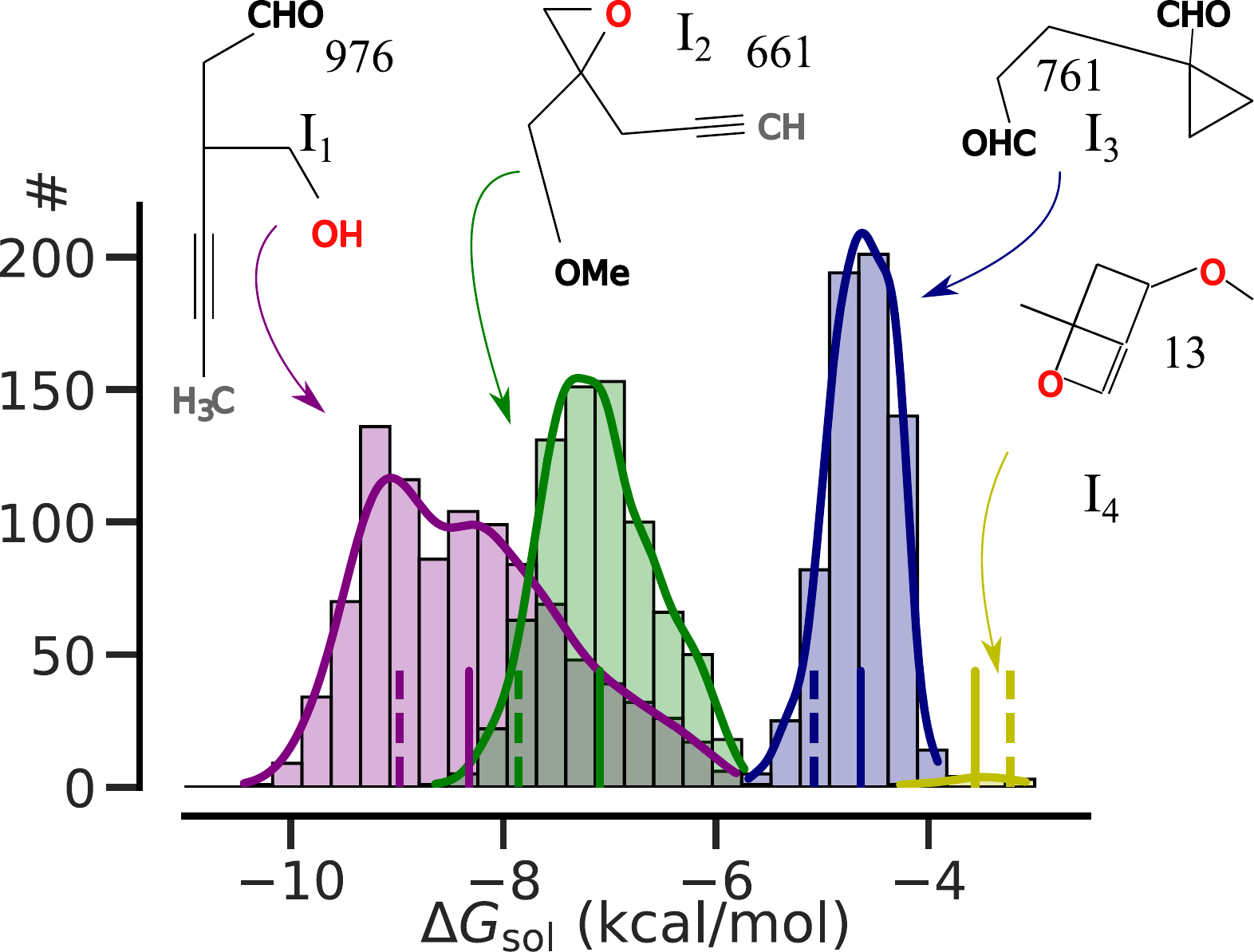}
    \caption{
        Distribution of hydration free energies of conformers at 298K
        estimated by QML for four constitutional isomers ($\text{C}_{7}\text{H}_{10}\text{O}_2$).
        Corresponding FML estimates, as well as averages, are denoted by solid and dashed vertical lines, respectively.
        Insets show molecular graphs of corresponding constitutional isomers and number of conformers.}
    \label{fig:dome_conformers}
\end{figure}

We randomly select four isomers of $\text{C}_{7}\text{H}_{10}\text{O}_2$ and use their complete set of conformers resulting from systematic scanning of all dihedral angles of the isomers\cite{clockwork} to compare the FML and QML predictions.
The distributions of the corresponding QML predictions of the free energy are shown in Fig.~\ref{fig:dome_conformers}. They reveal that for different configurations of the same isomer $\text{I}$, the predicted free energy $\Delta G_{\text{sol}}$ can vary by several kcal/mol, well above the experimental uncertainty.
Since hundreds of conformational isomers may exist for any given medium sized constitutional isomer
(the number of conformational isomers shown in Fig.~\ref{fig:dome_conformers} is $\text{I}_1, \dots, \text{I}_4 =  976,~ 661, ~761, ~13~$)
it should be obvious that using a single geometry as the representation may lead to considerable prediction errors.

Interestingly, we find (s. sec.~\ref{sec:results}) that due to an error cancellation QML based on vacuum geometries with weight coefficients $\boldsymbol{\alpha}^{\text{vac}}$ can reach mean absolute errors (MAEs) comparable with FML (s. sec.~\ref{sec:learning}).
The reason is that the average of the QML predictions over a set of conformers, denoted by $\langle . \rangle_{c}$ is close to the FML prediction,
\begin{align}
    \underbrace{
    \bigg  \langle
    \sum_{j}^{N^{\text{train}}} \alpha^{\text{QML}}_{j} K(\mathbf{X}^{\text{QML}}_{j},\mathbf{X}_{c} )  \bigg  \rangle_{c}
    }_{\text{Average QML}}
    \approx
    \underbrace{
    \sum_{j}^{N^{\text{train}}}
    \alpha_{j}^{\text{FML}}
    K(\langle \mathbf{X}_{j} \rangle ,\langle \mathbf{X}_{c} \rangle_{c} )}_{\text{FML}}   ~,
    \label{eq:approximation}
\end{align}
using the average representation over all conformers $\langle \mathbf{X}_{c} \rangle_c$. This can be seen by inspecting Fig.~\ref{fig:dome_conformers}, where the FML prediction of the average representation of all conformers is shown as a dashed line and the average of the individual conformer QML free energies as a solid line. The FML weight coefficients $\boldsymbol{\alpha}^{\text{FML}}$ are obtained using the representation average $\langle \mathbf{X}_{j} \rangle$ (s. Eq.~\ref{eq:average_representation}) over $s^{\text{train}} = 300$ MD samples (s. Fig.~\ref{fig:workflowA}).

\subsection{$\Delta$-Machine Learning} \label{sec:deltaML}

Using $\Delta$-ML\cite{deltaML} we promote solvation model (sm) free energies $\Delta G_{\text{sol}}^{\text{sm}}$ to experimental (exp) uncertainty as follows,
\begin{align}
    \Delta_{\text{sm}}^{\text{exp}} (\langle \mathbf{X} \rangle) \approx \Delta G_{\text{sol}}^{\text{exp}} - \Delta G_{\text{sol}}^{\text{sm}}~,
    \label{eq:deltaDifference}
\end{align}
using the ensemble representation $\langle \mathbf{X} \rangle$ as defined in Eq.~\ref{eq:average_representation}.
Note that the FML correction is different for each compound accounting for underestimations as well as overestimations of baseline free energy models. The step for promoting free energies to a predicted experimental value $\Delta G_{\text{sol}}^{\text{exp}}(\text{FML})$ is given by,
\begin{align}
    \Delta G_{\text{sol}}^{\text{exp}}(\langle \mathbf{X} \rangle) \approx
    \Delta G_{\text{sol}}^{\text{sm}} + \Delta_{\text{sm}}^{\text{exp}} (\langle \mathbf{X} \rangle)~.
    \label{eq:deltaML}
\end{align}
Besides these modifications the workflow remains the same as in Fig.~\ref{fig:workflowA}.

\section{Computational Details}  \label{sec:comput}

\subsection{Molecular Dynamics}

All MD simulations are performed with OpenMM\cite{openmm} in vacuum in the NVT ensemble using a Langevin integrator with a friction coefficient of $\SI{1}{\per \pico \second}$ and the small molecule FF GAFF2\cite{amber1, amber2} with a time-step of $\Delta t = \SI{2}{\femto \second }$ and total simulation time of $\SI{2}{\nano \second}$ using SHAKE\cite{shake}.
Partial charges are computed with antechamber\cite{amber1,amber2} at AM1-BCC\cite{bcc} level. An exception are the implicit solvent GBSA\cite{born, sasa} simulations where we use AMBER\cite{amber1}/GAFF\cite{amber2} while all other simulation parameters are the same. MD samples were selected with $\SI{2}{\pico \second}$ separation which is well beyond the maximal correlation time of $\mathtt{\sim} \SI{0.5}{\pico \second}$.

\subsection{Machine Learning}

To optimize the hyper-parameters $\sigma$ and $\lambda$ (s. Eqns.~\ref{eq:gauss}, \ref{eq:solution}), we perform 10-fold cross validation on the training set only and validate the performance on the test set using the QML package \cite{qmlcode}.
The training set contains all FreeSolv\cite{FreeSolv} molecules except for those 146 test set molecules which are also part of the QM9 database\cite{qm9}.
\\
Histograms showing the free energy distribution and the size of all molecules in the FreeSolv database are shown in the SI. For numerous FreeSolv compounds no value for the experimental uncertainty is reported, but a default value of
0.6 kcal/mol corresponding to thermal energy fluctuations at $\SI{298.15}{\kelvin}$ is assumed which we use as our target accuracy, albeit the total average absolute error of all experimental values is slightly smaller (0.57 kcal/mol).

\section{Results} \label{sec:results}

\subsection{Learning Curves and Free Energies of 116k Molecules}

\label{sec:learning}
\begin{figure*}
    \centering
    \includegraphics[width=\linewidth]{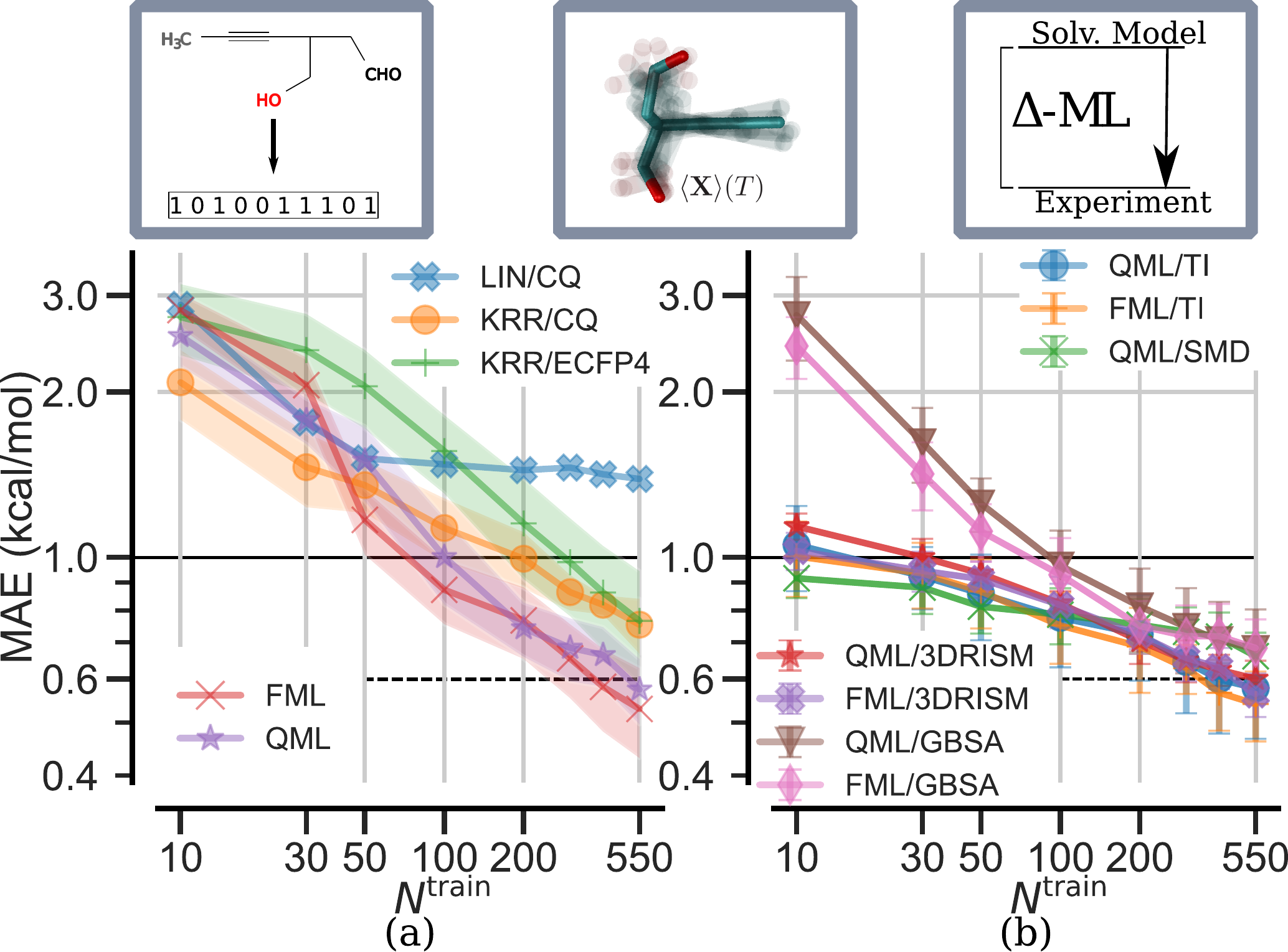}
    \caption{
        Learning curves for free energies of FreeSolv database\cite{FreeSolv} including standard deviation at each number of training molecules $N$ obtained by 10-fold cross-validation for feature based KRR, QML and FML in (a). QML/FML based $\Delta$-ML with various solvation models as baseline in (b). Solid black line marks thermo-chemical accuracy at 1 kcal/mol, dashed line the experimental uncertainty ($\SI{298}{\kelvin} \cdot  k_{B} \approx \SI{0.6}{\kilo \cal \per \mol}$).
        The boxes above illustrate the different ML approaches, in the first approach the molecule is featurized (as in ECFP4\cite{ecfp}) but all 3d information is lost. In the second case (Boltzmann weighted) 3d structures of the molecule are included (FCHL19\cite{FCHL, felixFCHL}). Thirdly, physics based approaches can be used as a baseline for $\Delta$-ML\cite{deltaML} to improve ML predictions, as in (b).
    }
    \label{fig:learning_curves}
\end{figure*}

The performance of several ML models is assessed using learning curves i.e. the prediction error is reported as a function of training set size $N$ (s. Fig.~\ref{fig:learning_curves}). Most notably we assess FML, the ensemble average representation (s. Eq.~\ref{eq:average_representation}) and $\Delta$-ML (s. Eq.~\ref{eq:deltaML}) for predicting experimental free energies of the FreeSolv\cite{FreeSolv} database.

In addition we also test the RDKit\cite{rdkit} implementation of the extended connectivity fingerprint\cite{ecfp} (ECFP4) commonly used in cheminformatics as well as a custom QSAR based representation, named CQ (s. SI). To compare KRR with conventional fitting methods we also report the MAEs of a multilinear regression model with CQ.
The error bars and colored areas in Fig.~\ref{fig:learning_curves} show the standard deviation of the MAEs resulting from 10-fold cross validation over the FreeSolv database.
We find that feature based models ECFP4 and CQ (s. Fig.~\ref{fig:learning_curves}a) perform worst reaching an MAE of only 0.8 kcal/mol for the maximal training set size of $N^{\text{train}} = 550$. This might be due to the fact that feature based representations do not properly weigh the ensemble of molecular conformations which can have a large influence on predictions of $\Delta G_{\text{sol}}$ (s. sec.~\ref{sec:krr_theory}).
\\

QML on the other hand reaches the target accuracy, the experimentally relevant accuracy of 0.6 kcal/mol for about $N^{\text{train}}= 550$ training molecules. The best FML model trained with 10 random MD samples per molecule with $T=\SI{350}{\kelvin}$ essentially results in the same MAE. While both models reach the experimentally relevant accuracy, we find that FML has a slightly smaller offset hitting 1 kcal/mol for $N^{\text{train}} \approx 50 $ while the QML model trained on the vacuum geometries has an MAE of about 1.5 kcal/mol. The otherwise similar performance of both models may to some extent be attributed to error compensation, as discussed above.

Learning curves for $\Delta$-ML\cite{deltaML} are also shown in Fig.~\ref{fig:learning_curves}, using various solvation models as a baseline such as GBSA (GAFF\cite{amber1, amber2}), 3D-RISM\cite{doi:10.1021/jp971083h, KOVALENKO1998237} (GAFF\cite{amber1, amber2}/TIP3P\cite{water}), SMD\cite{smd} (M06-2X\cite{functional}/Def2-TZVPP\cite{basisset}) and TI (GAFF2\cite{amber1, amber2}/TIP3P\cite{water}).
The GBSA values have been computed using AMBER\cite{amber1, amber2} (s. sec.~\ref{sec:comput}), the TI values are from the FreeSolv\cite{FreeSolv} database, the values for the two implicit solvation methods 3D-RISM\cite{doi:10.1021/jp971083h, KOVALENKO1998237} and SMD\cite{smd} are from the SI of \cite{validationset}.
We find that $\Delta$-ML can lower the offset of the learning curves and thus may be useful if experimental data is scarce.
FML requires about 100 training molecules to reach 0.8 kcal/mol whereas $\Delta$-ML using TI or 3D-RISM only needs around 50. Note that the SMD baseline model has the lowest offset, since SMD has the smallest MAE of 0.93 kcal/mol versus an MAE of 1.12 kcal/mol for TI on the FreeSolv database. However, after inclusion of $\sim$150 compounds in training both approaches, direct FML and $\Delta$-ML, perform similarly and converge to target accuracy.

Similar observations were also reported in Ref.~\cite{hybridML}, but we find that the target accuracy can be reached using FML without additional solvation calculations for $\Delta$-ML.
We note that some caution is required for this comparison since the accuracy of $\Delta$-ML relies heavily on systematic correlation between baseline and target.

\begin{figure*}
    \centering
    \includegraphics[width=0.7\textwidth]{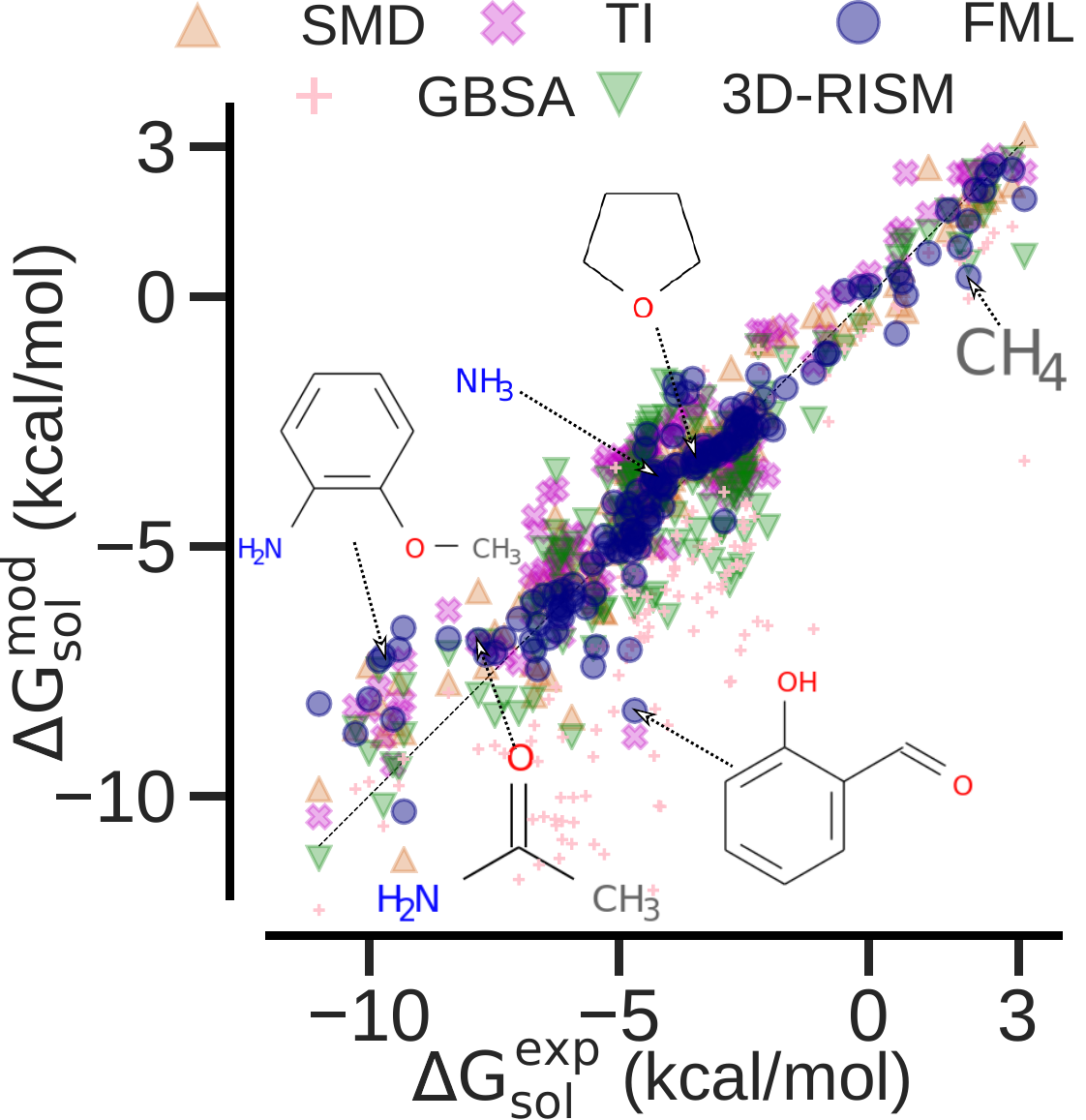}
    \caption{
        Scatter plot comparing free energies of solvation predicted by FML($\SI{350}{\kelvin}$; 10) (blue) using an ensemble based representation and solvation models: SMD (chocolate), TI (violet), 3D-RISM (green) and GBSA (pink). The FML predictions are also listed in the SI.}
    \label{fig:scatter_plot}
\end{figure*}

\subsection{Comparison to other models} \label{sec:models}

A comparison of various solvation models with FML is shown in terms of a scatter plot (s. Fig.~\ref{fig:scatter_plot}) displaying predictions versus experiment. The training set consists of all molecules in FreeSolv but not in QM9, the overlap (146 molecules) is used as a test set. The MAEs for FML and the tested solvation models are listed in Tab.~\ref{tab:MAE}. We find that the best FML model with $s^{\text{train}} = 10$ and $T^{\text{train}} = \SI{350}{\kelvin}$ results in an MAE of 0.57 kcal/mol reaching the experimentally relevant accuracy of solvation energy measurements and performing slightly better than SMD with 0.61 kcal/mol. Note that FML has some significant outliers such as methane with an error of about 2 kcal/mol. The largest outlier however is 2-hydroxybenzaldehyde with an error of 3.6 kcal/mol. It is worth emphasizing that SMD uses at least 79 of the test set molecules for parameterization which were not included in training of the FML model. Still FML has a slightly smaller MAE despite using fewer training points compared to SMD. Furthermore we find that FML clearly outperforms TI and 3D-RISM reaching an MAE of 1 kcal/mol after about $N^{\text{train}}=100$ training molecules. We emphasize that a typical SMD (M06-2X\cite{functional}/Def2-TZVPP\cite{basisset}) calculation of a molecule in the FreeSolv database with Gaussian09\cite{g16} is computationally demanding while a force-field based MD simulation in vacuum to generate the averaged representation can be run on a single modern central processing unit (CPU) in less than 10 minutes (s. Tab.~\ref{tab:MAE} for timing benchmarks).

In addition FML clearly outperforms COSMO\cite{cosmo, cosmo2, TURBOMOLE} with B-P86\cite{bp86,assesment}/TZVP\cite{doi:10.1063/1.467146} resulting in an MAE of 1.94 kcal/mol with higher computational costs. For comparison with COSMO-RS as implemented in COSMOtherm\cite{therm} we refer to a benchmark test\cite{doi:10.1021/ar800187p, KLAMT200043} with a set of 274 molecules resulting in an MAE of 0.52 kcal/mol reaching the accuracy of FML but at higher computational cost (s. Tab.~\ref{tab:MAE}) and as for all other
density functional theory (DFT) based methods with a worse scaling with the number of atoms compared to FFs. The predictions of FML for the test set are provided as a table in the SI.

To provide further insight in the prediction errors 642 FML models were trained on all molecules except for a single test molecule following a “one-versus-all” strategy.

This results in a mean unsigned error of 0.51 kcal/mol which is slightly smaller than the average reported experimental uncertainty of 0.57 kcal/mol. Note that for 450 of 642 FreeSolv\cite{FreeSolv} molecules a default error of $k_{B} \cdot T \approx \SI{0.6}{\kilo \cal \per \mol}$ was reported and our observation indicates that the “true” experimental error may actually be much smaller. Indeed we find that the average of the reported experimental uncertainties (other than default) is 0.48 kcal/mol. Given the slope and the offset of a continued learning curve (s.  Fig.~\ref{fig:learning_curves} in SI) as well as 0.48 kcal/mol as an upper boundary of the noise level, we can extrapolate that the learning curve would begin to flatten off for more than $N^{\text{train}}=670$ training molecules.
\\
We find that the two molecules with identifier $\text{mobley}\_2523689$ and $\text{mobley}\_3201701$ with the largest prediction errors of 5.3 kcal/mol and $\text{mobley}\_3201701$ 4.4 kcal/mol also have the largest reported experimental uncertainty of 1.93 kcal/mol. This may indicate that the reported experimental uncertainty may cause the large deviation for these outliers (s. Fig.~6 in SI).

Such a one-versus-all scheme may be useful to identify possible candidates with high experimental error, but we note that a large prediction error is not a sufficient condition but rather an indication of inconsistency with the rest of the data set. A large prediction error may also mean that the molecule in question has a very different chemical structure such that the structure-property relationship cannot be sufficiently explained by the training set.

\begin{table*}
    \caption{
        Comparison of MAEs, Pearson's $r$, estimated order of CPU time per solute prediction for various solvation models and FML. Approximate number of training molecules $N^{\text{train}}$ needed for FML to reach the MAE of each respective method. The conversion factor from GPU to CPU $c = \frac{t_{\text{CPU}}}{t_{\text{GPU}}}$ may vary substantially between $\mathtt{\sim}$~10 - 60 depending on hardware/MD code (here OpenMM\cite{openmm}). $g$ is the number of grid points for TI, typically $\mathtt{\sim}$~10.}
    \begin{tabular}{l|lSccccc}
        \hline
        \hline
        Class                & Model     & \text{MAE~(kcal/mol)}            & $r$  & $N^{\text{train}}$ & $\mathtt{\sim}~$CPU·h/solute & Reference                                                        & Year of Publication \\
        \toprule
        \multirow{4}{*}{DFT} & SMD       & 0.61\cite{validationset}         & 0.96 & 400                & \textit{high}                & [\citenum{g16, smd}~]                                            & 2009                \\
                             & COSMO     & 1.94                             & 0.90 & 20                 & $10^{-1}$                    & [\citenum{cosmo,cosmo2,TURBOMOLE}~]                              & 1993                \\
                             & COSMO-RS  & 0.52\cite{doi:10.1021/jp511158y} & 0.91 & --                 & $10^{-1}$                    & [\citenum{KLAMT200043, therm,doi:10.1021/jp511158y, TURBOMOLE}~] & 2000                \\
                             & DCOSMO-RS & 0.94\cite{doi:10.1021/jp511158y} & 0.87 & 100                & $10^{-1}$                    & [\citenum{dcosmo}~]                                              & 2006                \\
        \hline
        \multirow{3}{*}{FF}  & 3D-RISM   & 0.99\cite{validationset}         & 0.90 & 100                & $ 10^{-1}$                   & [\citenum{doi:10.1021/jp971083h, KOVALENKO1998237}]              & 1998                \\
                             & TI        & 0.93\cite{FreeSolv}              & 0.94 & 100                & $g \times c \times 10^{-1}$  & [\citenum{FreeSolv}~]                                            & 2017                \\
                             & GBSA      & 2.41                             & 0.84 & 20                 & $ 10^{-2}$                   & [\citenum{born, sasa}~]                                          & 2004                \\
        \hline
        \hline
        ML                   & FML       & 0.57                             & 0.95 & 490                & $10^{-2}$                    & this work                                                        & 2020                \\
        \hline
        \hline
    \end{tabular}
    \label{tab:MAE}
\end{table*}

\begin{figure*}[htb]
    \centering
    \includegraphics[width=\linewidth]{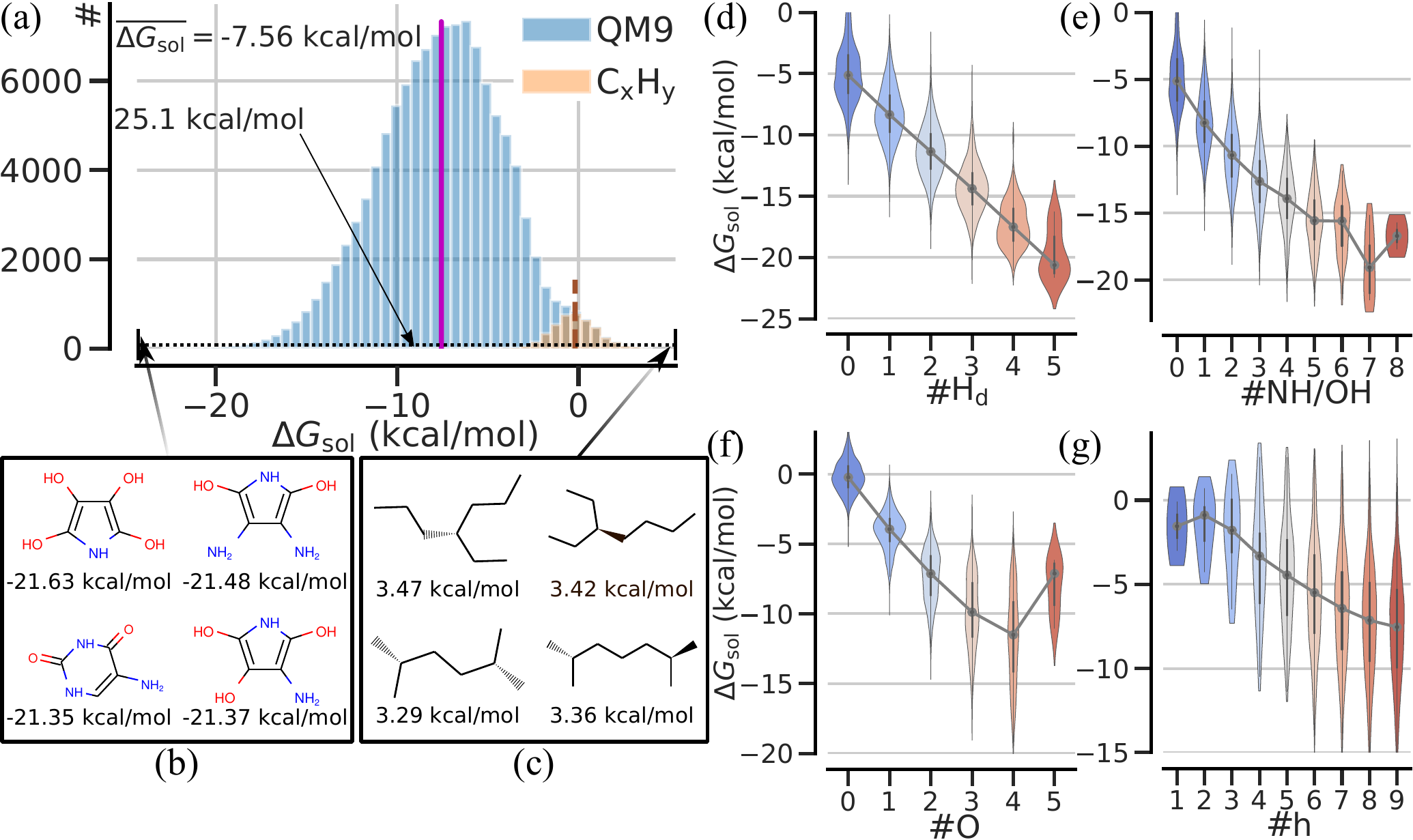}
    \caption{
    Free energy distribution of 116k molecules predicted by FML in (a), small histogram corresponds to a subset of 4907 hydrocarbons with averages shown as solid and dashed lines respectively. Below, four molecules with most negative (b) and positive (c) $\Delta G_{\text{sol}}$ in clockwise order.
    The mean of $\Delta G_{\text{sol}}$ for simple descriptors such as number of H-bond donors ${\rm H_{\rm d}}$ (d) NH/OH groups (e) or oxygen atoms $\rm O$ in stoichiometry $\text{C}_{n_{\text{C}}}\text{H}_{n_{\text{H}}}\text{O}_{n_{\text{O}}}$ (f) heavy atoms $\rm h$ (g).}
    \label{fig:qm9_histogram}
\end{figure*}

\subsection{Predicted solvation for 116k organic molecules} \label{sec:chemical_space}

We use FML to calculate free energies of a large dataset of organic molecules, a subset of CCS, in order to identify trends between solubility and structure.
More specifically, 116k molecules of the QM9 dataset have been considered in order to predict the free energy distribution in Fig.~\ref{fig:qm9_histogram}a.
For comparison we also show the corresponding distribution of $ \Delta G_{\text{sol}}$ for the FreeSolv database in the SI.
The FML model was trained on the complete FreeSolv database with coincident QM9 molecules removed from training. We find that the free energy distribution is approximately Gaussian with a mean value of $\overline{ \Delta G_{\text{sol}}}  = $ -7.56 kcal/mol spanning a range of 25.1 kcal/mol.
The molecules with the most negative and positive $\Delta G_{\text{sol}}$ are QM9 compounds indexed {\tt 26712} and {\tt 118570} with -21.63 kcal/mol and 3.47 kcal/mol, respectively (s. Fig.~\ref{fig:qm9_histogram}b,c).
The top 50 most soluble and least soluble molecules, according to FML estimates, are also shown in the SI.
We find that the most soluble molecules have a planar ring structure.
On the other hand the 4907 hydrocarbons of QM9 occupy the right tail of the distribution. Molecules with many hetero atoms bonded in NH/OH groups in a planar ring structure tend to have a very negative $\Delta G_{\text{sol}}$ while aliphatic linear molecules tend to be less soluble. This can be explained by the large difference in electronegativity between the H and the N, O atoms leading to polar bonds eventually resulting in H-bonds, lowering the enthalpy of solvation. Alkanes on the other hand have a negligible polarity, and thus only interact with water through much weaker van der Waals interaction resulting in low solubility.

These simple rules for the solubility lead to trends for the average free energy $\overline{ \Delta G_{\text{sol}}}$ such as approximately linear relations with the number of H-bond donors $\text{H}_{\text{d}}$ or the number of OH and NH groups $\text{NH}/\text{OH}$ as shown in Fig.~\ref{fig:qm9_histogram}de. Furthermore $\overline{ \Delta G_{\text{sol}}}$ decreases with the number of oxygen atoms for molecules with stoichiometry $\text{C}_{n_{\text{C}}}\text{H}_{n_{\text{H}}}\text{O}_{n_{\text{O}}}$ and with the number of heavy atoms $\text{h}$ (s. Fig.~\ref{fig:qm9_histogram}fg). The former effect is due to enthalpic H-bond contributions from hydroxyl groups and the latter is due to weak van der Waals interactions which roughly scale with the molecule size. Such linear free-energy relationships (LFER) are well known and used for predictive models e.g. for solvation \cite{linearRel, C8CP07562J} or organic chemistry\cite{D0SC04235H}.

The 116k predictions of FML for the subset of QM9\cite{qm9} are available in the SI.

\subsection{Analysis of the Model} \label{sec:analysis}

We study convergence of the MAE as a function of the number of MD samples for the training $s^{\text{train}}$ and test molecules $s^\text{query}$ given an MD sampling temperature of $T^{\text{query}} = T^{\text{train}} = \SI{350}{\kelvin}$. Note that both $s^\text{train}$ and $s^\text{query}$ refer to the representation (sum in Eq.~\ref{eq:average_representation}) and not to the number of training or test molecules (here always $620$ and $22$ resp. corresponding to 29 fold cross-validation over FreeSolv\cite{FreeSolv}).

\begin{figure}[htb]
    \centering
    \includegraphics[width=\columnwidth]{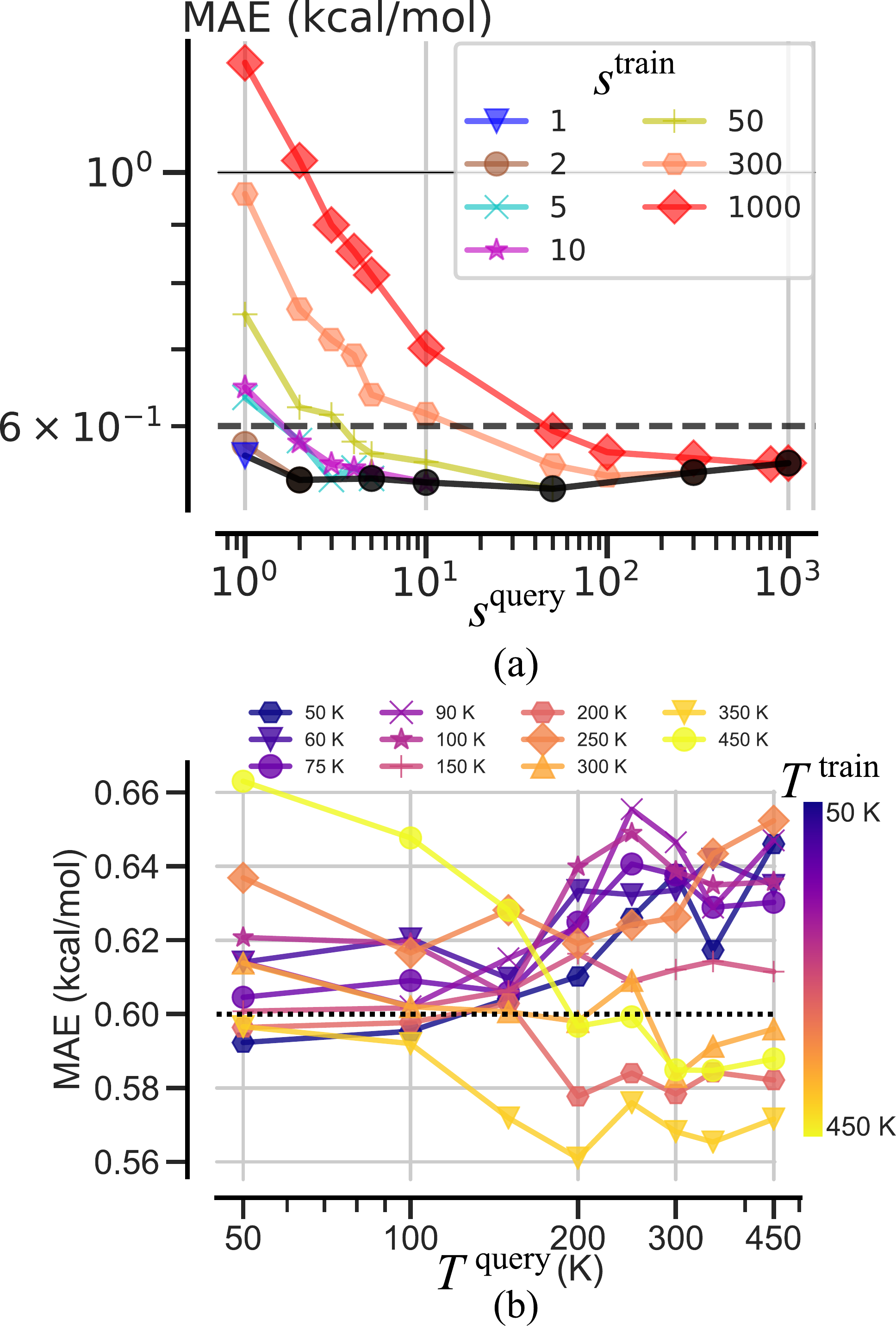}
    \caption{
        Comparison of MAEs of FML models using different number of samples in the training and query molecule representation ($T^{\text{query}} = T^{\text{train}} = \SI{350}{\kelvin}$) with $N^{\text{train}} =620$ distinct training molecules (a). Various temperatures for sampling the test $T^{\text{query}}$ and training molecules $T^{\text{train}}$ in (b) and resulting MAEs. Solid line marks thermo-chemical accuracy at 1 kcal/mol, dashed line the experimental uncertainty ($\SI{298}{\kelvin} \cdot  k_{B} \approx \SI{0.6}{\kilo \cal \per \mol}$) with $N^{\text{train}} =496$ training molecules evaluated on the test set with $146$ molecules.
    }
    \label{fig:convergence_nsamples}
\end{figure}

By inspection of Fig.~\ref{fig:convergence_nsamples}a we find that FML models trained with large $s^\text{train}$ tend to have a larger offset in the MAEs and a larger number of test samples $s^\text{query}$ is needed to reach the same accuracy as a model with small $s^\text{train}$. Note that we only show up to $s^\text{query} \leq s^\text{train} $ for the query molecule representation on the x-axis since a model evaluated on query FML representations with more MD samples than used for training would certainly not be desirable for ML.
All ML models improve with $s^{\text{query}}$ until the MAEs saturate at $s^\text{train} = s^\text{query} $ where FML generally achieves the highest accuracy. Interestingly, we find that the accuracy of the FML models increases only up to $s^\text{train} \approx 10$ and saturates beyond. This is because we find the FML representation to be sufficiently converged for approximately 10 random MD samples per molecule (s. Fig.~\ref{fig:convergence_rep}).

Next we investigate the temperature dependence of the FML representation. To this end the MAEs of several ML models $\text{FML}(T^{\text{train}}; ~ s^\text{train})$ with training molecule sampling temperatures $T^{\text{train}}$ between $\SI{50}{\kelvin}$ and $\SI{450}{\kelvin}$ are compared for various query molecule sampling temperatures $T^{\text{query}}$ with $s^\text{train} = s^\text{query} = 10$ (s. Fig.~\ref{fig:convergence_nsamples}b). We find that FML models usually perform best if $T^{\text{train}} \approx T^{\text{query}}$, e.g. model $\text{FML}(\SI{50}{\kelvin})$ performs best for $T^{\text{train}} = T^{\text{query}} = \SI{50}{\kelvin}$ and the MAE increases slightly for higher temperatures $T^{\text{query}}$. On the other hand, $\text{FML}(\SI{450}{\kelvin})$ performs worse at lower than at higher temperature. This indicates, as suggested by Eq.~\ref{eq:average_representation}, that the average representation indeed results in models specific for the temperature and phase space $\mathbf{\Gamma}(T)$ spanned by the training MD samples.

\section{Conclusion} \label{sec:conclusion}

We have studied the role of the representation for ML models of ensemble properties. As one would expect, numerical results confirm that the representation should be rooted in statistical mechanics, since ML based on a single molecular geometry can lead to large and spurious prediction errors. The definition of a representation based on Boltzmann weighted averages does not only resolve this issue but also naturally introduces temperature dependence.

In similar manner FML models could be constructed that depend on pressure, or chemical potential, to account for increasingly more realistic canonical and grand-canonical ensembles.
The numerical performance of FML is encouraging: FML reaches experimental uncertainty for relatively small training sets and at low computational cost for new query estimates.
As such, it is better or on par with state of the art models in the field, and it emerges as a viable alternative whenever sufficient training data is available.
Furthermore, we find that $\Delta$-ML can improve the predictions for small training set sizes, however FML can reach experimental relevant accuracy without requiring additional solvation model calculations. Furthermore we stress that it is straightforward to improve the accuracy and transferability of FML by adding more experimental data points. This is not necessarily true for other more conventional solvation models.
To demonstrate the usefulness of a transferable FML model, we have predicted solvation energies for 116k organic molecules of the QM9 database (see Fig.~\ref{fig:qm9_histogram}). The results confirm known trends namely that molecules with a high solubility tend to have many hetero atoms and are arranged in planar ring structures while linear aliphatic molecules tend to have a lower solubility. We have therefore demonstrated that FML can be used to study solvation throughout CCS, and that it might prove useful to identify additional non-trivial structure-property relationships.

\section{Supplementary Material}
See supplementary material for a table of predicted free energies of solvation on the test set and information about the CQ-representation. There we also show the predicted most and least soluble molecules of QM9 as well as the distribution of heavy atoms in the FreeSolv\cite{FreeSolv} database. In addition we show a histogram illustrating the convergence of the FML representation and the individual prediction errors and outliers of the model as well as an extended learning curve.

\section*{Acknowledgement}
O.A.v.L. acknowledges support from the Swiss National Science foundation (407540\_167186 NFP 75 Big Data) and from the European Research Council (ERC-CoG grant QML and H2020 projects BIG-MAP and TREX).
This project has received funding from the European Union's Horizon 2020 research and
innovation programme under Grant Agreements \#952165
and \#957189.
This project has received funding from the European Research Council (ERC) under the European Union’s Horizon 2020 research and innovation programme (grant agreement No. 772834).
This result only reflects the
author's view and the EU is not responsible for any use that may be made of the
information it contains.
This work was partly supported by the NCCR MARVEL, funded by the Swiss National Science Foundation.

\section{Data Availability Statement}

The 116k predictions of FML for the subset of QM9\cite{qm9} are available as an ASCII text file (QM9\_116k\_dG.txt) for download in the SI. The predictions of FML for the validation set are provided as a table in a PDF file in the SI (SI.pdf). All other data that support the findings of this study are available from the corresponding author upon reasonable request.

\bibliography{literature}
\end{document}